\documentclass[12pt,a4paper, amsmath,amssymb]{article}
\usepackage[margin=0.8in]{geometry}
\usepackage{amssymb,amsmath}
\usepackage{amsfonts}
\usepackage{bm}
\usepackage[utf8]{inputenc}
\usepackage{graphicx}
\usepackage{float}
\usepackage{amsmath}
\usepackage[colorlinks,linkcolor = blue,
urlcolor  = blue,
citecolor = blue,
anchorcolor = blue]{hyperref}
\usepackage{xcolor}
\usepackage{ctable} 
\usepackage{adjustbox}
\def\thefootnote{\fnsymbol{footnote}}
\usepackage{cite}

\begin{document}
	{
\begin{center}
{\Large \textbf{Analytical Critical Phenomena of Rotating Regular AdS Black Holes with Dark Energy }}
\thispagestyle{empty}
\vspace{1cm}

{\sc
	
	H. Laassiri$^1$\footnote{\url{hayat.laassiri@gmail.com}},
	A. Daassou$^1$\footnote{\url{ahmed.daassou@uca.ma}},
	R. Benbrik$^1$\footnote{\url{r.benbrik@uca.ac.ma}}\\
}
\vspace{1cm}
{\sl
	$^1$Fundamental and Applied Physics Laboratory, Physics Department, Polydisciplinary Faculty,
	Cadi Ayyad University, Sidi Bouzid, B.P. 4162, Safi, Morocco.\\
}
\end{center}
\vspace*{0.1cm}
\begin{abstract}
This study  focuses on precisely calculating analytical critical points for rotating regular AdS black holes, examining scenarios with and without external dark field contributions. Importantly,    it represents the first attempt to compute critical points for this specific class of rotating black holes. Our primary focus is on investigating the impact resulting from variations in the charge of nonlinear electrodynamics on the critical phenomena of rotating regular  AdS black holes,  while also incorporating the influence of quintessence field contributions. The analytical investigation is concentrated on the horizon radius, employing two distinct approaches to simplify the complexity and length of the calculations. Furthermore, our examination extends to deciphering the intricate relationship between dark energy and critical phenomena.  This involves visually portraying a range of critical behaviors  and detailing a recent discovery regarding how the intensity of quintessence affects phase transitions. The shift  in these transitions conform to either a concave or convex function, a characteristic dependent on the sign of quintessence intensity.

\end{abstract}

\def\thefootnote{\arabic{footnote}}
\setcounter{page}{0}
\setcounter{footnote}{0}

\newpage
\section{Introduction}
\label{intro}\
The existence of singularities signifies the  inability to predict specific physical  phenomena, underscoring the inadequacy of general relativity. Fortunately, attempts have been undertaken to overcome singularities, including the incorporation of regular black holes. Bardeen's creation of a regular black hole~\cite{1}  demonstrates this approach, incorporating ideas from the theory of quantum gravity and substituting central singularities with matter backgrounds. As the Bardeen solution did not satisfy the vacuum Einstein equations, Beato and Garcia demonstrated  that by considering the coupling of gravitational theory with nonlinear electrodynamics, one could derive this solution~\cite{2}. The exploration of regular solutions has been extensively pursued in the literature, showcasing a multitude of proposed solutions and comprehensive analyses of their properties~\cite{3,4,5,6,7}. This reflects a substantial body of work dedicated to understanding and expanding upon the implications of incorporating nonlinear electrodynamics in gravitational theory. \

There are similar aspects between non-linear
	electrodynamics and Hayward-Bardeen black holes. To begin,  both involve non-linear
	effects. In electrodynamics, this is evident in the form of non-linear equations describing
	the electromagnetic field, while in the context of Hayward-Bardeen black holes, non-
	linearity arises from modifications to the Einstein field equations due to the presence of
	exotic matter or scalar fields~\cite{8,9}
	Moreover, both non-linear electrodynamics and Hayward-Bardeen black holes influence
	the interaction between gravitational and electromagnetic fields. Non-linear electrody-
	namics introduces corrections to the electromagnetic field equations, potentially affecting
	the behavior of charged particles in curved spacetime. Similarly, Hayward-Bardeen black
	holes modify the gravitational field around them, consequently impacting the behavior of
	electromagnetic fields in their vicinity~\cite{10}.

 Extensive discussions in~\cite{11,12,13,14,15,16,17} focused on  how the quintessential parameter  impacts  the quasinormal frequencies across various black hole spacetimes. In reference to~\cite{18}, the authors investigated the Bardeen solution, incorporating a cosmological constant and surrounded  it by quintessence. Their study revealed that this solution could be derived through the coupling of the Einstein equations with nonlinear electrodynamics. Importantly, the authors highlighted that the solution is not always regular, outlining the  necessary conditions for its regularity. Additionally, they conducted   an in-depth analysis  of the thermodynamics associated with this solution, establishing both the form of the Smarr formula and the first law of thermodynamics in this context. This comprehensive exploration  provides valuable insights into the interplay between quintessence, nonlinear electrodynamics, and the properties of black hole solutions. The examination of critical phenomena in the context of Anti-de Sitter (AdS) black holes has been  a focal point in  various of scholarly works, each making valuable contributions to the expanding comprehension of this intricate field~\cite{19,20,21,22,23}. Extensive  research is ongoing to  comprehend  phase transitions in quintessential black holes from  diverse perspectives. The   existence of a dark sector has  prompted  an examination  into the  thermodynamics concerning both charged and regular black holes, as explored in previous studies~\cite{24,25}. \

This research extensively explores the thermodynamic characteristics and critical phenomena exhibited by rotating regular AdS black holes within different theoretical frameworks. The paper is organized as follows:  In Section \ref{2}, our focus shifts to the examination of rotating Bardeen black holes. We conduct a detailed analysis to understand how  the charge  parameter in nonlinear electrodynamics affects critical phenomena. Furthermore, we employ a precise calculation method to accurately determine the critical point associated with these rotating black holes. Section \ref{3} is  dedicated  to computing and formulating the analytical expression for critical point in Kerr Hayward AdS black holes. It also involves analyzing the phase transitions of this particular category of AdS black holes.  We explore the impact of thermal fluctuations on the stability of Kerr-Hayward-AdS.
 Section \ref{4} delves into the exploration of black holes characterized by a small horizon radius. Within this section, our research is dedicated to formulating an approximate method for deriving analytical expressions of critical points specific to these black holes when subjected to the influence of dark energy. Moving to Section \ref{5}, our attention is directed toward studying the impact of dark energy on black holes with  a large horizon radius. Here, we utilize an approximate approach to establish analytical expressions for critical points, providing insights into the interplay between dark energy and black hole properties. In Section \ref{6}, we explore the influence of dark energy on the phase transitions of rotating Bardeen-AdS black holes. This involves a comprehensive analysis of various critical behaviors, contributing to a deeper understanding of the complex interactions at play. Finally, Section \ref{7} serves as a pivotal integration point, where we consolidate the outcomes obtained from the preceding sections. This structured approach allows for a systematic exploration of the diverse aspects of Bardeen-AdS black holes, enriching our understanding of their thermodynamic properties and critical behaviors across different scenarios.

\section{Unveiling  key  dynamics:  Critical  phenomena of  rotating  Bardeen-AdS  black  holes}
\label{2}
The derivation of the line element governing the rotation of Bardeen-AdS black holes is established in connection with a cosmological constant, as explained in  Ref.~\cite{26},\
\begin{align}
{\text{ds}^2=\dfrac{-\Delta _r}{\Sigma }\left(\text{dt}-\dfrac{\text{asin}^2\theta }{E}\text{d$\phi $}\right)^2+\dfrac{\Sigma }{\Delta
		_r}\text{dr}^2+\dfrac{\Sigma }{\Delta _{\theta }}\text{d$\theta $}^2+\dfrac{\Delta _{\theta }\sin ^2\theta }{\Sigma }\left(\text{adt}-\dfrac{r^2+a^2}{E}\text{d$\phi
		$}\right)^2},\label{Eq 1}
\end{align}

where 
\begin{equation}
 {\Delta_{r}=-2 m r \left(\frac{r^2}{g^2+r^2}\right)^{3/2}+\left(a^2+r^2\right) \left(1+\frac{8}{3} P \pi  r^2\right)}.
\end{equation}\label{Eq 2}
In this context, $ g $ symbolizes the charge of nonlinear electrodynamics, $ m $ stands for the mass parameter, and $ a $ represents the rotational parameters of the black hole,

with
\begin{align}\label{Eq 3}
{{r=\dfrac{\sqrt{3 S-8 a^2 P \pi  S-3 a^2 \pi }}{\sqrt{3 \pi } }}},\hspace{2cm}{j= \dfrac{a   m}{\Xi ^2}}, \hspace{2cm}{\Xi = \frac{3 - 8 a^2 P \pi}{3}}.
\end{align}

 Using the previously obtained expression for the horizon radius  derived from the entropy formula of a rotating Bardeen-AdS black hole, and incorporating it into the equation for angular momentum,  we obtain the following equation,
\begin{equation}
{j = \dfrac{a S (3+8 P S)\left(-3 \left(g^2 \pi +S\right)+a^2 \pi  (3+8 P S)\right)^{3/2}}{2 \sqrt{3 \pi } \sqrt{3 S-a^2 \pi  (3+8 P
			S)} \left(-3 S+a^2 \pi  (3+8 P S)\right)^{3/2}}}\label{Eq 4}.
\end{equation}
By using Eq. \ref{Eq 3} and  incorporating the derived expression for $a(P, S, g)$ from Eq. \ref{Eq 4} into the term $\dfrac{\Delta'_{r}}{4 \pi (a^{2}+ r^{2})}$,  we obtain the following expression representing the Hawking temperature,
\begin{align}\label{Eq 5}
{T=\frac{t_1+t_2+t_3}{4 B \sqrt{3 \pi } S \sqrt{3 B^2 S-A^2 \pi  (3+8 P S)} \left(-3 B^2 \left(g^2 \pi +S\right)+A^2 \pi  (3+8 P S)\right)}}.
\end{align}

Obtaining an analytical expression for the critical point involves solving a system of equations derived from the critical condition,
\[
\dfrac{\partial T}{\partial S}\bigg|_{P,g,j} = \dfrac{\partial^2 T}{\partial S^2}\bigg|_{P,g,j} = 0,
\]
which reduces to polynomial equations of degree greater than five with arbitrary coefficients in \(S\). This makes exact resolution impossible without resorting to approximations. In our approach, we used the Taylor series to express the angular momentum \(j\) and truncated the series at the second order to make the mathematical problem tractable. This approximation is necessary to effectively address the problem of determining the critical point.

After a rigorous calculation, we derive the analytical expression for the critical point of rotating Bardeen-AdS black holes as follows.
\begin{align}\label{Eq 6}
P_c&=\left(24 g^2 \left(179 g^4+435 j^2\right) \pi ^3 P_1+P_2+P_3-\left(648\ 2^{2/3} \left(17 g^4-30 j^2\right)^4 \pi ^8+P_4+P_5-P_6\right)
	\gamma _1\right.\notag\\
&\left.-3 \left(450\ 2^{1/3} \left(17 g^5-30 g j^2\right)^2 \pi ^5+P_7\right) \gamma _0^{1/3} \gamma _1^{3/2}\right)/\left(407040 g^{10} \pi
	^5 \gamma _0^{2/3} \gamma _1\right)\notag,\\
S_c&=\dfrac{1}{2} \left(9 g^2 \pi +\sqrt{\gamma _1}+\sqrt{\gamma _2}\right),
\end{align}
with \(\gamma_0\), \(\gamma_1\),  and \(\gamma_2\) represented in relation to \(j\) and \(g\) (see Appendix).
	
To verify the accuracy of the analytical approximations derived in this study, we substituted $ j=0 $ into our results and compared them with the outcomes obtained in  Ref.~\cite{27} for a non-rotating Bardeen-AdS black hole. Using relative deviation, we observed   $  \lvert \Delta T_{c} \rvert=0.35\% $ and  $  \lvert \Delta P_{c} \rvert=0.86\% $. This substantiates the robust validity of the approximate expressions obtained. \	
	
To evaluate the influence of the nonlinear electrodynamics charge on phase transitions, we generate graphs illustrating the relationship between temperature $ (T) $ and entropy $ (S) $, as well as Gibbs free energy $ (G) $ plotted against temperature $ (T) $. To carry out this analysis, we utilize Eq. \ref{Eq 5} and incorporate the provided expression for $ G $.\ 
\begin{equation}\label{Eq 7}	
{G=\dfrac{6 j^2 \pi ^2 S \left(4 g^2 \pi +S\right)+\left(g^2 \pi +S\right)^4 (3+8 P S)}{6 \sqrt{\pi } S \left(g^2 \pi +S\right)^{5/2}}-T S}.\
	\end{equation}
	\begin{figure}[H]
	\centering
	\includegraphics[width=0.45\linewidth, height=5.8cm]{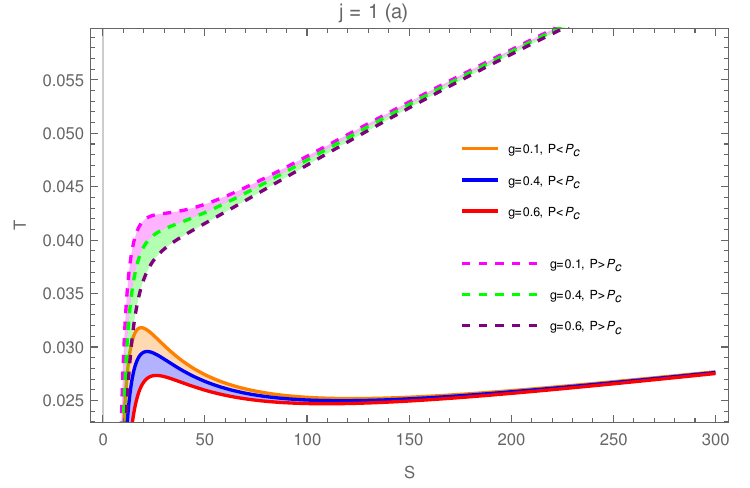}\hspace{1cm}	\includegraphics[width=0.45\linewidth, height=5.8cm]{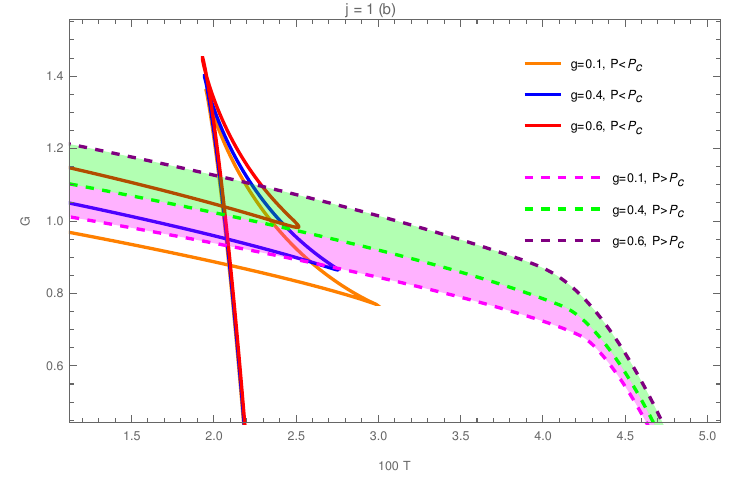} 
	
	\caption{Graph (a): Temperature vs. entropy and  graph (b): Gibbs  free  energy vs. temperature for  varied $ g $ in  rotating Bardeen-AdS black holes.}
	\label{fig1}
\end{figure}	
	Graph (a) portrays the fluctuation of temperature with respect to entropy for $ P < P_{c} $. The plot demonstrates oscillatory behavior that diminishes for $ P > P_{c} $. The extremum points shift towards higher temperature values as the nonlinear electromagnetic charge decreases. Graph (b) visually represents the changes in Gibbs free energy as a function of temperature. This distinctive swallowtail pattern becomes evident when the pressure $P $ is below the critical level, showcasing fluctuations. As the pressure surpasses the critical point, the swallowtail effect diminishes. Additionally, the introduction of a higher nonlinear electromagnetic charge $g$ not only influences the intersection point but also causes a shift in the minimal temperature $ T_{min} $ towards higher values, accompanied by an increase in Gibbs free energy.
\section{Analytical  aspects of  rotating Hayward-AdS black holes}
\label{3}
We  derived the thermodynamic quantities from the line element  outlined in  Ref.~\cite{28}. The expression for $ \Delta(r) $ is formulated in terms of $ a $, $ M $, and $ g_{h} $, where $ a $  a signifies the rotating parameter, $ M $ stands for the mass parameter, and $ g_{h} $  indicates the charge associated with non-linear electrodynamics,
\begin{align}\label{Eq 8}
{\Delta _{r} =\left(a^2+r^2\right) \left(1+\frac{8}{3} P \pi  r^2\right)-\frac{2 M r^4}{r^3+g_h^3}}.
\end{align}
By applying the formula  represented as $ \dfrac{\Delta'r}{4 \pi (a^{2}+ r^{2})} $, we express the representation of the Hawking temperature  associated with Hayward AdS black holes  as follow.
\begin{align}\label{Eq 9}
T=&r \left(3+8 a^2 P \pi +16 P \pi  r^2-\frac{2 \left(a^2+r^2\right) \left(3+8 P \pi  r^2\right)}{r^2}+\frac{3 r \left(a^2+r^2\right)
	\left(3+8 P \pi  r^2\right)}{2 \left(r^3+g_h^3\right)}\right)\notag\\
&\left/\left(6 \pi  \left(a^2+r^2\right)\right)\right..
\end{align}
In determining the entropy corrections due to thermal fluctuations, we base our calculations on the canonical partition function.
\begin{align}\label{Eq 10}
{Z(\beta ) = \int _0^{\infty }\sigma (E)\exp ^{-\text{$\beta $E}}\text{dE}},
\end{align}
the symbol \( \sigma(E) \) denotes the quantum density of the system. By applying the inverse Laplace transformation, we derive the density of states,
\begin{align}\label{Eq 11}
\sigma (E)= \frac{-i}{2 \pi}\int _{c-\text{i$\infty $}}^{c+\text{i$\infty $}}Z(\beta )\exp ^{\text{$\beta $E}}\text{d$\beta
	$},
\end{align}
with $ S_{0} $ denoting the corrected entropy, it is represented as,
\begin{align}\label{Eq 12}
{S_0=\text{$\beta $E}+\text{Log}[Z]}.
\end{align}
After thorough and rigorous calculation, the formulation of the density of states is derived as follows,	
\begin{align}\label{Eq 13}
{\sigma (E) = \dfrac{-i e^S}{2\pi}\int _{c-\text{i$\infty $}}^{c+\text{i$\infty $}}\exp ^{\dfrac{(\beta -c)^2}{2}\dfrac{d^2S_0}{d^2\beta
	}}\text{d$\beta $}}.
\end{align}
After simplifying the calculations, the corrected entropy is expressed as a function of temperature, entropy, and corrected parameters,
\begin{align}\label{Eq 14}
{S_0=-\beta \text{Log}\left(T^2 S \right)+\frac{\beta _0}{S}+S}.
\end{align}
By substituting the content of Eq. \ref{Eq 9}  into Eq. \ref{Eq 14}, we formulate a modified expression. Setting the value of $ \beta_{0} $ to 0 yields  the resulting expression,
\begin{align}\label{Eq 15}
S_0&=-\beta  \text{Log}\left(-\frac{\left(a^2 r^3 \left(-3+8 P \pi  r^2\right)+3 \left(r^5+8 P \pi  r^7\right)-2 \left(3 r^2+a^2 \left(6+8
	P \pi  r^2\right)\right) g_h^3\right){}^2}{48 \pi  \left(-3+8 a^2 P \pi \right) r^2 \left(a^2+r^2\right) \left(r^3+g_h^3\right){}^2}\right)\notag\\
&+\frac{3 \pi  \left(a^2+r^2\right)}{3-8 a^2 P \pi}.
\end{align}
\begin{figure}[H]
	\centering
	\includegraphics[width=0.48\linewidth, height=6.7cm]{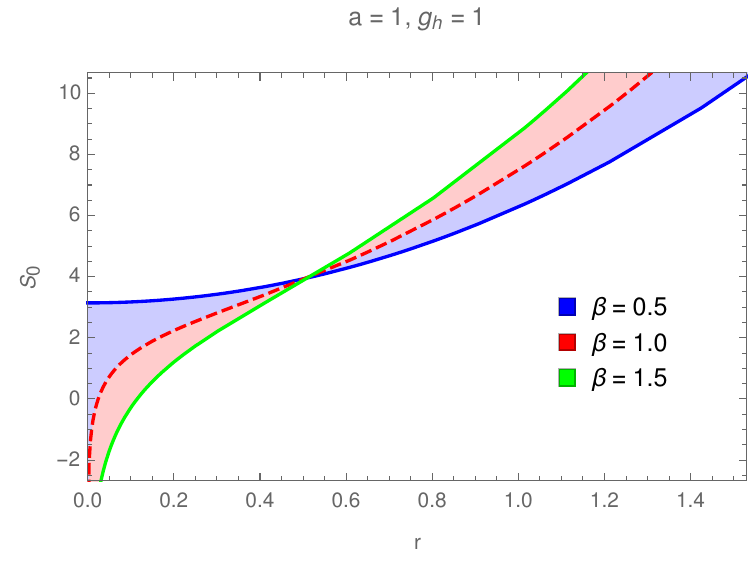}
	\caption{Dependence of  corrected  entropy on $ r $ for  different  values of  corrected  parameter.}
	\label{fig2}
\end{figure}
The graphical representation delineates the fluctuations in both corrected and uncorrected entropy. Particularly noteworthy is the consistent observation that at equilibrium ($ \beta = 0 $), entropy not only retains a positive value but also exhibits a continuous and incremental rise. This trend aligns seamlessly with the underlying principles of the second law of black hole thermodynamics, which stipulates a perpetual increase in black hole entropy. Nevertheless, as the correction parameter $ \beta $ diverges from zero, smaller black holes tend to display a proclivity for negative entropy. This tendency becomes notably more pronounced with elevated $ \beta $ values.  Additionally, for larger horizon radius, the corrected entropy consistently takes on positive values, resembling the pattern seen in uncorrected entropy. This discovery holds noteworthy implications, highlighting the considerable influence of thermal fluctuations on the thermodynamics of smaller black holes.\

To derive and  analytically express the critical point of the Kerr Hayward-AdS black hole, we follow a specific methodology outlined in two distinct cases. The first case pertains to a simplified scenario where the angular momentum parameter $ (a) $ is set to zero, and the second case deals with a more complex situation where the angular momentum parameter differs from zero.\

For the case where $ a = 0 $, the Hawking temperature is formulated as follows:
\begin{align}\label{Eq 16}
{T=\frac{r^3+8 P \pi  r^5-2 g_h^3}{4 \pi  r^4+4 \pi  r g_h^3}}.
\end{align}
By applying the critical condition  $ {\dfrac{\partial T}{\partial r}|_{j, g_{h}, P}}=
{	\dfrac{\partial ^2T}{\partial ^2r}|_{j, g_{h}, P}} $, the analytical expression for the critical point can be expressed as follows:\
\begin{align}\label{Eq 17}
&{P_c=\dfrac{3 \left(57-23 \sqrt{6}\right) \left(14+6 \sqrt{6}\right)^{1/3}}{800 \pi { g_h^2{}}}}\notag,\\
&{v_c=2\left(14+6 \sqrt{6}\right)^{1/3} {g_h{}}}\notag,\\
&{T_c=\dfrac{\left(5-2 \sqrt{6}\right) \left(7+3 \sqrt{6}\right)^{2/3}}{4\ 2^{1/3} \pi  { g_h{}}}}.
\end{align}
We discover that the equation of state suggests a universal ratio, given by:
\[
\rho_{h} =\dfrac{P_cv_c}{T_c}=\dfrac{-3\sqrt{6}+27}{50}\simeq 0.393,
\]
this ratio differs from the universal one identified for Van der Waals fluids, where $ \rho=0.375 $.

In the scenario where angular momentum is non-zero, in the small limit of $j$, we obtain the expressions for $r$ and $a$ from the equations of entropy ($S$) and angular momentum ($j$) of the Hayward-AdS black hole  as  indicated below,
\begin{align}\label{Eq 18}
{{r=\frac{\sqrt{3-8 a^2 P \pi } \sqrt{3 a^2 \pi -3 S+8 a^2 P \pi  S}}{\sqrt{3 \pi } \sqrt{-3+8 a^2 P \pi }}}},
\end{align}
\begin{align}\label{Eq 19}
{a=\dfrac{6 j \sqrt{\pi } S}{(3+8 P S) \left(S^{3/2}+\pi ^{3/2} g_h^3\right)}+O\left(j^2\right)}.
\end{align}
By replacing the expressions from Eq. \ref{Eq 18} and Eq. \ref{Eq 19} into Eq. \ref{Eq 9}, we derive.
\\
\begin{align}\label{Eq 20}
T=&\left(-2 \pi ^6 g_h^{12} \left(3+8 P S\right)+\pi ^{9/2} g_h^9 S^{3/2} \left(-15-16 P S+64 P^2 S^2\right)+S^4
\left(3+8 P S\right) \right.\notag\\
&\left(-6 j^2 \pi ^2+S^2+8 P S^3\right)+3 \pi ^3 g_h^6 S \left(-12 j^2 \pi ^2+S^2 \left(-3+16 P S+64 P^2 S^2\right)\right)+\notag\\
&\left.\pi ^{3/2} g_h^3 S^{5/2} \left(-12 j^2 \pi ^2 \left(9+16 P S\right)+S^2 \left(3+80 P S+192 P^2 S^2\right)\right)\right)/\left(4
\sqrt{\pi } \sqrt{S} \right.\notag\\
&\left.\left(3+8 P S\right) \left(\pi ^{3/2} g_h^3+S^{3/2}\right){}^4\right).
\end{align}

By applying the critical condition $ {\dfrac{\partial T}{\partial S}|_{j, g_{h}, P}}=
{	\dfrac{\partial ^2T}{\partial ^2S}|_{j, g_{h}, P}} $,  and adopting  the same procedure used in Section 2, after simplifying, we obtain the following derivation.
\begin{align}\label{Eq 21}
&P_c=\left(\xi _1+c_2 \left(\xi _4-\xi _5-\xi _6\right)+18 c_1^{2/3} c_2 \left(\xi _7+\xi _8\right)+3 c_1^{1/3}
\sqrt{c_2} \left(10 \pi ^3 g_h^6 \left(\xi _2+\xi _3\right)\right.\right.+3 c_2\notag\\
&\left.\left. \left(\xi _9+\xi _{10}\right)\right)\right)/\xi _{11}\notag,\\
&S_c=\frac{1}{2} \left(\sqrt{c_2}+\sqrt{c_3}\right),
\end{align}
 with \hspace{1cm}  $\xi_1, \xi_2, \xi_3, \ldots, \xi_{10}$ and $c_1, c_2, c_3$ are expressed in terms of $g_h$ and $j$ (see Appendix).

To analyze the phase transition of Kerr-Hayward AdS black holes, we utilize Eq. \ref{Eq 20} to plot $ T $ as a function of $ S $. Additionally, we employ the formula $ C = T \dfrac{dS}{dT}  $ to graph the heat capacity as a function of entropy $ S $. 

 The expression for Gibbs energy is determined by employing the equation $ G = m - TS $, along with Eq. \ref{Eq 8} and Eq. \ref{Eq 20}, where m denotes the physical mass of Kerr-Hayward  -Ads black holes. The mass parameter is established by imposing the condition $ \Delta(r) = 0 $.

\begin{align} \label{Eq 22}
G=&-T S+\left(\pi ^{9/2} g_h^9 \left(3+8 P S\right)+3 \pi ^3 g_h^6 S^{3/2} \left(3+8 P S\right)+S^{5/2} \left(6 j^2
\pi ^2+ \left(3+8 P S\right)\right)\right.\notag\\
&\left.S^2+3 \pi ^{3/2} g_h^3 S \left(8 j^2 \pi ^2+S^2 \left(3+8 P S\right)\right)\right)/\left(6 \sqrt{\pi } S \left(\pi ^{3/2} g_h^3+S^{3/2}\right){}^2\right).
\end{align}

\begin{figure}[H]
	\centering
	\includegraphics[width=0.38\linewidth, height=5.8cm]{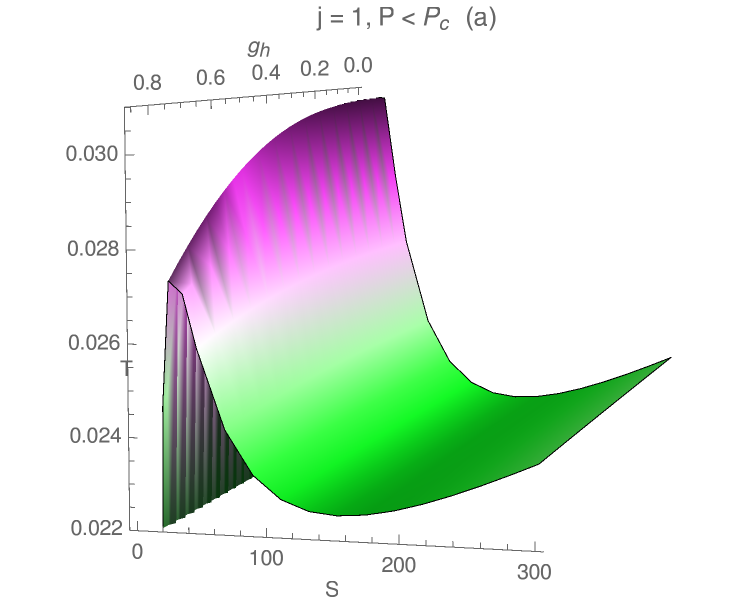}\hspace{0.01cm}		\includegraphics[width=0.4\linewidth, height=5.8cm]{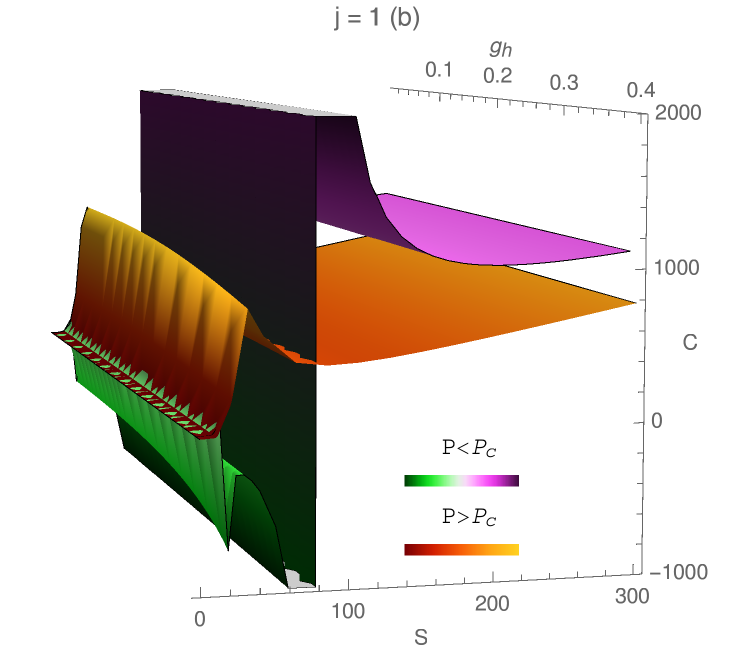}
	
	\caption{ Graph (a): Temperature vs. entropy, Graph (b): Heat  capacity vs. entropy.}
	\label{fig3}
\end{figure}
\begin{figure}[H]
	\centering
	\includegraphics[width=0.38\linewidth, height=5.8cm]{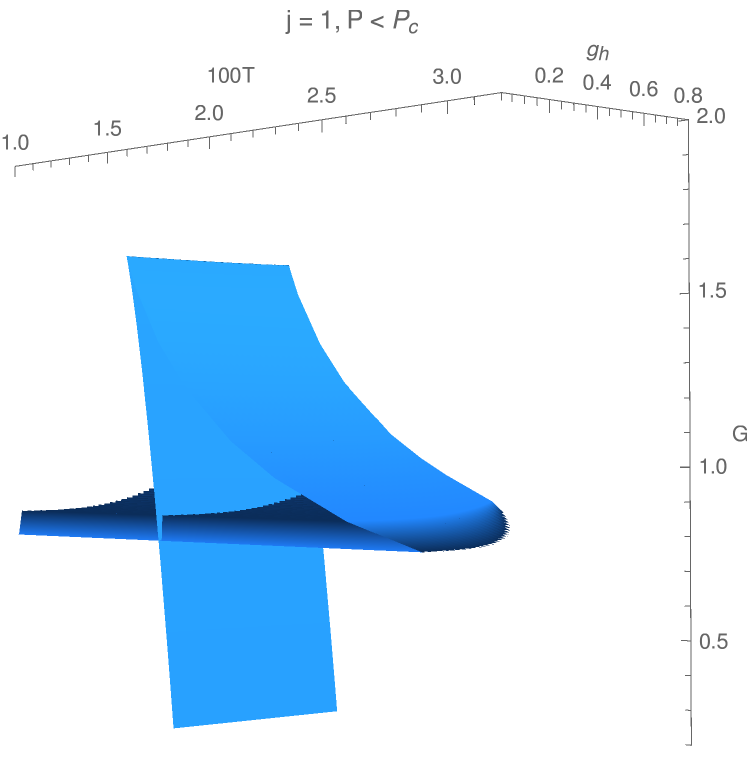}
	
	\caption{  Gibbs  energy vs. temperature for  non-linear  electrodynamics  charge.}
	\label{fig4}
\end{figure}
The description of the temperature-entropy  plot (Figure. \ref{fig3}, graph (a)) under conditions where pressure $P$ is below the critical pressure shows oscillatory behavior.  Two extreme points  move towards higher temperatures as the charge of the non-linear electrodynamics decreases,  creating a concave curve with $g_{h} = 0$ symbolizing the peak. For graph (b): Below the critical pressure, a negative slope indicates an intermediate black hole branch with negative heat capacity, suggesting system instability. Heat capacity divergence signals an ongoing phase transition. Above the critical pressure, no phase transition occurs, restoring system stability. The Gibbs diagram (Figure. \ref{fig4}) reveals a distinct swallowtail pattern, noticeable when the pressure is less than the critical pressure. As the charge of the non-linear electrodynamics rises, the intersection point shifts towards higher $G$ and lower $T$.

\begin{table}[htbp]
	\centering

	\begin{tabular}{|c|c|c|c|c|}
		\hline
		\multicolumn{5}{|c|}{\textbf{rotating Bardeen-AdS black holes}} \\
		\hline
		\textbf{$g$} & \textbf{$ P_{c} $} & \textbf{$ S_{c} $} & \textbf{$ T_{c} $} & \textbf{$ G_{c} $} \\
		\hline
		0.3 & 0.010547 & 6.069231 & 0.076726 & 0.405439 \\
		0.5 & 0.004516 & 12.931001 & 0.049630 & 0.639051 \\
		0.8 & 0.001822 & 31.279590 & 0.031444 & 1.012064 \\
		\hline
		\multicolumn{5}{|c|}{\textbf{rotating Hayward-AdS black holes}} \\
		\hline
		\textbf{$g_{h}$} & \textbf{$ P_{c} $} & \textbf{$ S_{c} $} & \textbf{$ T_{c} $} & \textbf{$ G_{c} $} \\
		\hline
		0.3 & 0.013261 & 6.186845 &  0.091029 &  0.304038 \\
		0.5 & 0.006796 & 12.029185 & 0.064658 & 0.435181 \\
		0.8 & 0.003441 & 23.736345 & 0.045291 & 0.640423 \\
		\hline
	\end{tabular}
	\caption{Critical thermodynamic quantities for rotating Bardeen and Hayward-AdS black holes.}\label{bb}
\end{table}
	
Table \ref{bb} presents numerical results for various critical thermodynamic quantities for rotating Bardeen-AdS black holes and rotating Hayward-AdS black holes. The aim is to compare the effect of the non-linear electrodynamics charge on these properties. It has been observed that the non-linear electrodynamics charge exerts a similar influence on both rotating Bardeen-AdS and rotating Hayward-AdS black holes. An increase in the charge values results in a decrease in the critical temperature and critical pressure. Conversely, the critical entropy and Gibbs energy increase in value, indicating that a higher non-linear electrodynamics charge leads to a lower phase transition temperature and a larger associated entropy.	
\section{Analytical  expression for  critical  point in  small  horizon  radius of  rotating Bardeen  AdS  black  holes with  dark  energy}	
\label{4}
We derive the thermodynamic properties of rotating Bardeen-AdS black holes that include a quintessence field, referencing~\cite{11}, and ~\cite{29}. Our primary objective is to investigate the influence of dark energy on the critical phenomena exhibited by these black holes. This section is specifically dedicated to exploring the dynamics of rotating  Bardeen-AdS black holes enriched with dark energy, with a particular emphasis on situations where the horizon radius is characterized by small values.\
\begin{align}
{\Delta _r=-2 m r \left(\dfrac{r^2}{b^2+r^2}\right)^{3/2}+\left(a^2+r^2\right) \left(1+\dfrac{8}{3} P \pi  r^2\right)-r^{-1-3 \omega
	} \alpha }\label{Eq 23},
\end{align}
 in this context, $b$  is indicative of the nonlinear electromagnetic charge, while $ \omega $ represents the state parameter for quintessence. Additionally, $ \alpha $ is used to characterize the  intensity of the quintessence.\

 Deriving an analytical expression for the critical point involves solving a set of equations derived from the critical condition, \noindent\({\dfrac{\partial T}{\partial S}|_{P, b, j, \omega, \alpha}=\dfrac{\partial ^2T}{\partial ^2S}|_{P, j, b, \omega, \alpha}=0}\), which relies on the Hawking temperature derived from $\dfrac{\Delta'_{r}}{4 \pi (a^{2}+ r^{2})}$. These equations are of arbitrary order due to the presence of the $ -1-3\omega $ term in the expression for $ \Delta_{r} $, making exact resolution unfeasible without employing approximations. To tackle this mathematical challenge, we used the Taylor series to represent the horizon, as depicted in Eq.\ref {Eq 23}.  This approximation was necessary to effectively address the task of determining the critical point. Truncating the series at the first order provided results comparable to those of rotating Bardeen-AdS black holes, when we set additional parameters to zero. Moreover, restricting the order to 1 helps avoid unnecessary complexities.
\begin{align}
{r^{-1-3 \omega }=2+3 \omega -r (1+3 \omega )+O(r-1)^2}\label{Eq 24},
\end{align}
\begin{align}
{m= \dfrac{\left(b^2+r^2\right) \left(3 a^2+3 r^2+8 a^2 P \pi  r^2+8 P \pi  r^4-6 \alpha +3 r \alpha -9 \alpha  \omega +9 r \alpha  \omega \right)}{6
		r^3 \sqrt{\dfrac{r^2}{b^2+r^2}}}},\label{Eq 25}
\end{align}	
 the mass parameter $ m $ is deduced from Eq. \ref{Eq 23} when $\Delta_{r}=0$, and this derivation involves  applying the Taylor expansion to Eq. \ref{Eq 24}, which  describes the horizon radius.
\begin{align}
 T&=\dfrac{-a^2 \left(3 r^2-8 P \pi  r^4+4 b^2 \left(3+4 P \pi  r^2\right)\right)+3 \left(b^2 \left(-2 r^2+4 \alpha  (2+3 \omega )-3
		r (\alpha +3 \alpha  \omega )\right)\right)}{12 \pi  r \left(a^2+r^2\right) \left(b^2+r^2\right) } \notag \\	
&{ +\dfrac{3r^2 \left(r^2+8 P \pi  r^4+\alpha  (2+3 \omega )\right)}{12 \pi  r \left(a^2+r^2\right) \left(b^2+r^2\right)} }.\label{Eq 26}
\end{align}
 In the small limit of $j$, we represent the parameter of angular momentum $ a$ as,
\begin{align}
{a=\dfrac{6 j \sqrt{\pi } S^{3/2} \sqrt{\dfrac{S}{b^2 \pi +S}}}{\left(b^2 \pi +S\right) \left(3 S+8 P S^2-6 \pi  \alpha +3 \sqrt{\pi
		} \sqrt{S} \alpha -9 \pi  \alpha  \omega +9 \sqrt{\pi } \sqrt{S} \alpha  \omega \right)}+O(j)^2}.\label{Eq 27}
\end{align}
Through the replacement of the expressions from Eq. \ref{Eq 3} and Eq. \ref{Eq 27} into Eq. \ref{Eq 25}, and subsequent application of the critical condition, we obtained an approximate analytical expression for the critical point through a sequence of simplifications.\  
\begin{align}	
&{P_c=\dfrac{\left(P_{1S}+
		P_{2S}+
		P_{3S}+
		P_{4S}\right)}{\left(8480 b^{10} \pi ^5\right)}}\notag,\\
&{S_c=\dfrac{1}{4} \left(18 b^2 \pi -12 \pi  \alpha -18 \pi  \alpha  \omega \right)+\dfrac{\sqrt{\lambda _4}}{2}+\dfrac{\sqrt{\lambda
			_5}}{2}},\
\end{align}	\label{Eq 28}	
 with $P_{1s}$, $P_{2S}$, $P_{3S}$, $P_{_{4S}}$, $\lambda_{4}$, $\lambda_{5}$ expressed in terms of $b$, $j$, $\omega$, $\alpha$ (see the appendix).

 Setting the quintessential state parameter $ \omega $ = 0 and the intensity of quintessence $ \alpha $ = 0 yields identical results to those obtained in Section \ref{2}.		
\section{Deriving  analytical  expression for  critical  point of  rotating Bardeen-AdS  black  holes with  dark  energy at  large  horizon  radius}
\label{5}
In this particular section, our attention is directed towards analyzing how quintessence affects large rotating Bardeen-AdS black holes. This involves deriving the analytical expression for the critical point.  We employ the same procedure as in Section \ref{4}  To facilitate this analysis, we adopt the following approximation, with $ n $ is the convergence value of the event horizon radius.\
\begin{align}
r^{-1-3 \omega }={n^{-2-3 \omega } (-r (1+3 \omega )+n (2+3 \omega ))+O(r-n)^2}.\label{Eq 29}
\end{align}
The earlier equation is derived by employing the Taylor series expansion of the horizon radius expression, which converges towards values much larger than 1. We limited the Taylor expansion to the first order for the same reasons discussed in Section \ref{4},
\begin{align}
{m=\dfrac{n^{-2-3 \omega } \left(a^2 n^{2+3 \omega } \left(3+8 P \pi  r^2\right)+n^{2+3 \omega } r^2 \left(3+8 P \pi  r^2\right)+3
		r \alpha  (1+3 \omega )-3 n \alpha  (2+3 \omega )\right)}{6 r \left(\dfrac{r^2}{b^2+r^2}\right)^{3/2}}},\label{Eq 30}
	\end{align}
the mass parameter of a rotating Bardeen-AdS black hole with dark energy, particularly for large values of the horizon radius, is derived through the utilization of a specific Eq. \ref{Eq 23}.	
	\begin{align}
	T &=\left(-a^2 n^{2+3 \omega } \left(3 r^2-8 P \pi  r^4+4 b^2 \left(3+4 P \pi  r^2\right)\right)+3 \left(n r^2 \left(n^{1+3 \omega
		} \left(r^2+8 P \pi  r^4\right)+\alpha  (2+3 \omega )\right)\right.\right.\notag\\
&\left.\left.+b^2 \left(-2 n^{2+3 \omega } r^2-3 r \alpha  (1+3 \omega )+4 n \alpha  (2+3 \omega )\right)\right)\right)/\left(12 \pi  r \left(a^2+r^2\right)
		\left(b^2+r^2\right)n^{-2-3 \omega } \right)\label{Eq 31}.
	\end{align}
In order to determine the critical point, the angular momentum parameter must be expressed in relation to $ P $ and $ S $ as follows,
	\begin{align}
	{a=\dfrac{6 j n^{2+3 \omega } \sqrt{\pi } \sqrt{S} \left(\dfrac{S}{b^2 \pi +S}\right)^{3/2}}{n^{2+3 \omega } S (3+8 P S)+3 \sqrt{\pi
			} \sqrt{S} \alpha  (1+3 \omega )-3 n \pi  \alpha  (2+3 \omega )}+O(j)^2}.\label{Eq 32}
		\end{align}
		Substituting Eq. \ref{Eq 3} and Eq. \ref{Eq 32} into Eq. \ref{Eq 31} while adopting the identical critical condition from Section \ref{4}, we provide approximate analytical expression for the critical point of rotating Bardeen-Ads black holes enveloped by quintessence. This analysis is specifically tailored for scenarios characterized by significantly large values of the horizon radius.\
		\begin{align}\label{Eq 33}
		&	{P_c= \dfrac{n^{-1-3 \omega }(P_{1L}+P_{2L}+
			P_{3L})}{8480
			b^{10} \pi ^5}},\notag\\		
&{S_c=\dfrac{3}{2} n^{-1-3 \omega } \pi  \left(3 b^2 n^{1+3 \omega }-2 \alpha -3 \alpha  \omega \right)+\dfrac{\sqrt{\beta _3}}{2}+\dfrac{\sqrt{\beta
				_4}}{2}},	\
	\end{align}		
with	$ \beta_{1} $, $ \beta_{2} $, $ \beta_{3} $, and $ \beta_{4} $ are formulated in relation to $ j $, $ n $, $ g $, $ \omega $, $ b $, and $ \alpha $ (see Appendix).

				By setting the external parameters of the quintessence field $ \alpha $ = 0  and  $ \omega $ = 0, we observe consistent results with those detailed in Section \ref{2}.
				
	\begin{table}[htbp]
		\centering
		\label{tab:thermo_properties}
		\begin{tabular}{|c|c|c|c|c|c|}
			\hline
			\textbf{$\omega$} & \textbf{$b$}  & \textbf{$P_{c}$} & \textbf{$ S_{c} $} & \textbf{$T_{{c}} $} & \textbf{{$ G_{c} $}} \\
			\hline
			-1 & 0.3 & 0.008177 & 7.707532 & 0.067932  & 0.373279\\
			& 0.5 & 0.003956 & 14.931977 & 0.046928  & 0.579808\\
			& 0.8 & 0.001720 & 33.432813 & 0.030801  & 0.931611\\
			\hline
			$\frac{-2}{3}$ & 0.3 & 0.010987 & 5.724030 & 0.078332  & 0.349819\\
			& 0.5 & 0.004639 & 12.345093 & 0.050434   & 0.577299\\
			& 0.8 & 0.001844 & 30.627317 & 0.031718  & 0.949997 \\
			\hline
			$\frac{-1}{3}$ & 0.3 & 0.017056 & 3.645610 & 0.096235  & 0.308704 \\
			& 0.5 & 0.005760 & 9.18768 & 0.055334   & 0.564254\\
			& 0.8 & 0.001996 & 27.566450 & 0.032755   & 0.966299 \\
			\hline
		\end{tabular}
		\caption{Critical thermodynamic quantities for different values of $\omega$ and $b$ with $j = \alpha = 0.1  $.}\label{b1}
	\end{table}	
  The calculations are presented in Table \ref{b1}. It has been observed that, for different values of the equation of state parameter $\omega$, an increase in the charge of nonlinear electrodynamics results in a decrease in the critical temperature and pressure, while the critical entropy and Gibbs free energy increase. Furthermore, when the value of $b$ is fixed, it is noted that as $\omega q$ increases, the critical pressure and temperature increase, whereas the critical entropy and Gibbs free energy decrease.
		
				\newpage	
				\section{Critical  behaviors of  rotating  bardeen AdS  black  holes with  quintessence  field}
				\label{6}
				In this section, we use the  following expressions for temperature and Gibbs free energy. These expressions are general and applicable to both small and large values of the horizon radius for rotating Bardeen-AdS black holes with dark energy. We graph the relationship between temperature $T$ and entropy $S$, with a specific emphasis on scenarios where the pressure is  below  the critical pressure.\
				\begin{align}
			{T}= \dfrac{S^{-2-\frac{3 \omega }{2}} \left(12 j^2 \pi ^2 S^{4+3 \omega } \left(b^2 \pi +S\right) \left(-S^2 T_2+b^2 \pi  S T_3+b^4
				T_4\right)+T_1 T_5\right)}{4 \sqrt{\pi } \left(b^2 \pi +S\right) T_5}.\label{Eq 34}
						\end{align}
	\( T_{1}, T_{2}, T_{3}, T_{4} ,T_{5}\) are expressed in terms of \( j \), \( b \), \( \alpha \), \( \omega \), \( P \), and \( S \).  Further details can be found in the Appendix.

	\begin{figure}[H]
		\centering
		\includegraphics[width=0.4\linewidth, height=5.5cm]{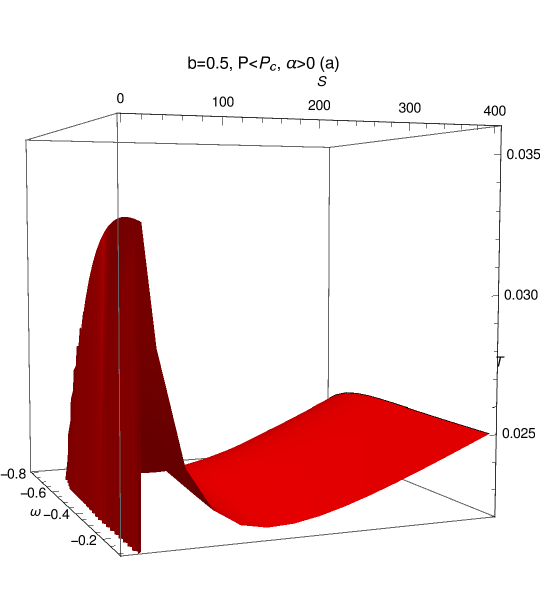}\hspace{1cm}	\includegraphics[width=0.4\linewidth, height=5.5cm]{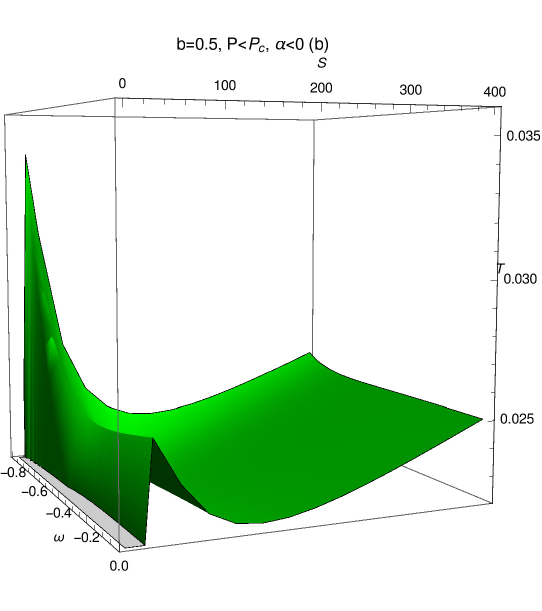} 
		
		\caption{ Exploring the  impact of  quintessence  intensity on  temperature:  Positive values (graph a) and  negative  values (graph b). }
		\label{fig5}
	\end{figure}
The temperature-versus-entropy plot, considering diverse values of the quintessential state parameter $ \omega $, reveals oscillatory patterns for both positive and negative quintessence intensities when the pressure is below the critical point. Specifically, in instances of positive quintessence intensity, the extrema indicate a decreasing trend in temperature as $ \omega $ decreases, shaping a concave function. Conversely, for negative quintessence intensity, the extrema show an increase as $ \omega $ values decrease, illustrating a convex function.\
\begin{align}
{G=-S T+\frac{S^{-\frac{3}{2} (2+\omega )} \left(36 j^2 \pi ^2 S^{5+3 \omega } \left(S M_2+M_3\right)+\left(b^2 \pi +S\right)^3 M_1
		M_4\right)}{36 \sqrt{\pi } \sqrt{\frac{S}{b^2 \pi +S}} \left(b^2 \pi +S\right)^3 M_4}}.\label{Eq 35}	
	\end{align}	
 The expressions for \( M_{1} \), \( M_{2} \), \( M_{3} \), and \( M_{4} \) are presented using the variables \( j \), \( b \), \( \alpha \), \( \omega \), and \( S \). Further details can be found in the Appendix.	
			
To assess the impact of the quintessence field on critical behaviors, we represent the Gibbs free energy as a function of entropy in the subsequent plots.
	\begin{figure}[H]
	\centering
	\includegraphics[width=0.45\linewidth, height=5.8cm]{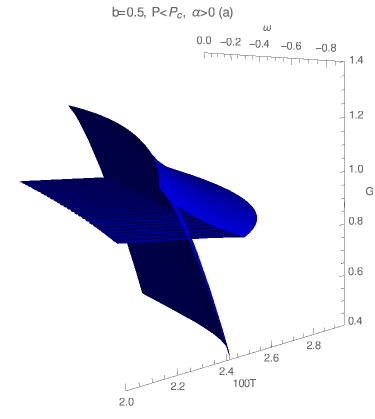}\hspace{1cm}	\includegraphics[width=0.45\linewidth, height=5.8cm]{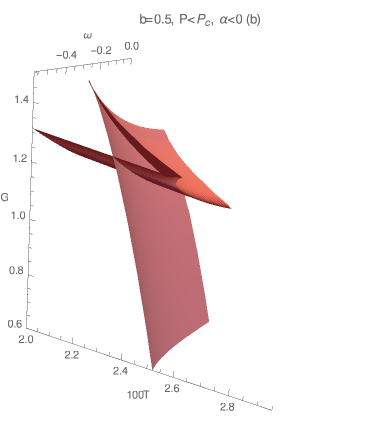} 
	
	\caption{ Gibbs  free  energy in  relation to  temperature for  various $ \omega $  values: Positive $ \alpha $ ( graph a) and  negative $ \alpha $  scenarios (  graph b).  }
	\label{fig6}
\end{figure}
The plotted graph illustrating Gibbs energy as a function of temperature, considering different $ \omega $ values and various quintessence intensity scenarios $ \alpha $, showcases distinctive swallowtail patterns when the pressure is below the critical threshold. Positive $ \alpha $ values lead to the intersection point and minimal temperature shifting towards higher temperature and Gibbs free energy values as $ \omega $ increases, delineating a concave pattern. Conversely, for negative $ \alpha $ values, the intersection point and minimal temperature shift towards lower Gibbs free energy and higher temperature values as $ \omega $ increases, revealing a convex pattern.			
\section{Conclusion}
\label{7}
In this recent research,  we focus centrally on the notable impact of dark energy on determining the critical point of rotating Bardeen-AdS black holes . We formulated approximate analytical expressions to establish a nuanced relationship between the parameters and thermodynamic features of rotating Bardeen-AdS black holes and those linked to the dark sector. Rigorous validation procedures were implemented to verify the accuracy and reliability of the derived expressions for critical point. This research offers a deeper understanding of the thermodynamic characteristics of rotating Hayward-AdS black holes, highlighting how thermal fluctuations affect both large and small black holes. The introduction of a non-zero correction parameter leads smaller black holes to favor negative entropy.  We derive  analytical expressions for critical point, exploring the impact of the non-linear electrodynamics charge on the phase transition of rotating Hayward-AdS black holes. Moreover, our examination  extends to unraveling the influence of the charge of nonlinear electrodynamics on the critical phenomena of rotating regular AdS black holes.\

By illustrating diverse critical behaviors, we  underscore the impact of quintessence intensity on the dynamics of phase transitions. Specifically, when quintessence intensity is positive, the extrema imply a decline in temperature as $ \omega $ decreases, forming a concave function. Conversely, under negative quintessence intensity, the extrema suggest an increase as $ \omega $ values decrease, depicting a convex function. Additionally, our findings revealed that positive $ \alpha $ values prompt the intersection point and minimal temperature to shift towards higher temperature and Gibbs free energy values as $ \omega $ increases, resulting in a concave pattern. Conversely, negative $ \alpha $ values cause the intersection point and minimal temperature to move towards lower Gibbs free energy and higher temperature values as $ \omega $ increases, unveiling a convex pattern. The novel findings emerging from this study underscore the substantial influence exerted by the quintessence field on the critical phenomena observed in rotating Bardeen-AdS black holes.

\section*{ Acknowledgements}
 The authors would like to thank the anonymous referee for interesting comments and suggestions which motivated us to prepare a well-improved revised version.

\newpage
\appendix

\textbf{\Large{Appendix}}\

  Appendix A: Rotating Bardeen-AdS black holes

\noindent\({t_1=-2 A^4 \pi ^2 (1+4 P S) (3+8 P S)^2}\),

\noindent\({t_2=-9 B^4 S \left(-2 g^2 \pi +S+8 P S^2\right)}\),

\noindent\({t_3=3 A^2 B^2 \pi  (3+8 P S) \left(2 g^2 \pi +S (3+16 P S)\right)}\),

\noindent\({A=6 j \sqrt{\pi } \left(\frac{S}{g^2 \pi +S}\right)^{3/2}}\),

\noindent\({B=\sqrt{S} (3+8 P S)}\),

\noindent\({P_1=\left(36\ 2^{1/3} \left(17 g^4-30 j^2\right)^2 \pi ^4+15 \left(37 g^4+48 j^2\right) \pi ^2 \gamma _0^{1/3}+2\ 2^{2/3} \gamma
	_0^{2/3}\right) \gamma _0^{1/3} \sqrt{\gamma _1}}\),\

\noindent\({P_2=6 \gamma _0^{2/3} \gamma _1^{5/2} \left(25 g^2 \pi +12 \sqrt{\gamma _2}\right)}\),\

\noindent\({P_3=288 g^4 \left(179 g^4+435 j^2\right)^2 \pi ^6 \gamma _0^{2/3}}\),\

\noindent\({P_4=4\ 2^{1/3} \gamma _0^{4/3}-270\ 2^{1/3} \left(17 g^4-30 j^2\right)^2 \pi ^5 \gamma _0^{1/3} \left(44 g^4 \pi +48 j^2 \pi +5 g^2
	\sqrt{\gamma _2}\right)}\),\

\noindent\({P_5=6 \gamma _0^{2/3} \left(12 \left(1691 g^8+9670 g^4 j^2+2700 j^4\right) \pi ^4+12 g^2 \left(1076 g^4+2115 j^2\right) \pi ^3 \sqrt{\gamma
		_2}-25 g^2 \pi  \gamma _2^{3/2}\right)}\),\

\noindent\({P_6=15\ 2^{2/3} \pi  \gamma _0 \left(44 g^4 \pi +48 j^2 \pi +5 g^2 \sqrt{\gamma _2}\right)}\),\

\noindent\({P_7=25\ 2^{2/3} g^2 \pi  \gamma _0^{2/3}+6 \gamma _0^{1/3} \left(29 g^2 \left(101 g^4+240 j^2\right) \pi ^3+10 \left(73 g^4+72 j^2\right)
	\pi ^2 \sqrt{\gamma _2}-4 \gamma _2^{3/2}\right)}\),\

$ \gamma _0 $ ,$ \gamma _1 $, $ \gamma _2 $ are represented in relation to 'j' and 'g' in the following expressions,

\noindent\({\gamma _0=54 \pi ^6\left(4737 g^{12}-18930 g^8 j^2+63900 g^4 j^4-27000 j^6+ 4 \sqrt{5}\right.}\\
{\left.\sqrt{-g^4 \left(21230 g^{20}-952893 g^{16} j^2+2047500 g^{12} j^4+275400 g^8 j^6-19926000 g^4 j^8+12150000 j^{10}\right)}\right) }\),\

\noindent\({\gamma _1=\dfrac{18\ 2^{1/3} \left(17 g^4-30 j^2\right)^2 \pi ^4+30 \left(11 g^4+12 j^2\right) \pi ^2 \gamma _0^{1/3}+2^{2/3} \gamma
		_0^{2/3}}{6 \gamma _0^{1/3}}}\),\

\noindent\({\gamma _2=\dfrac{\left(-18 2^{1/3} \left(17 g^4-30 j^2\right)^2 \pi ^4+60 \left(11 g^4+12 j^2\right) \pi ^2 \gamma _0^{1/3}-2^{2/3}
		\gamma _0^{2/3}\right) }{6 \gamma _0^{1/3} }+\dfrac{24 g^2 \left(179 g^4+435 j^2\right) \pi ^3 }{6\text{  }\sqrt{\gamma _1}}}\).\
	
  Appendix B: Rotating Hayward-AdS black holes

\noindent\({\xi _1=-28800 \pi ^6 c_1^{2/3} g_h^{12}-18 c_1^{2/3} c_2^{5/2} \left(16 \sqrt{c_3}+5 \pi ^{3/2} g_h^3\right)}\),

\noindent\({\xi _2=32\ 2^{2/3} c_1^{2/3}+2304\ 2^{1/3} \pi ^3 \left(15 j^2 \sqrt{\pi }+28 g_h^3\right){}^2}\),

\noindent\({\xi _3=3 c_1^{1/3} \left(3840 j^2 \pi ^2+8192 \pi ^{3/2} g_h^3+25 \pi ^3 g_h^6\right)}\),

\noindent\({\xi _4=16\ 2^{1/3} c_1^{4/3}+41472\ 2^{2/3} \pi ^6 \left(15 j^2 \sqrt{\pi }+28 g_h^3\right){}^4}\),

\noindent\({\xi _5=3\ 2^{2/3} \pi ^{3/2} c_1 \left(960 j^2 \sqrt{\pi }+\left(2048+15 \sqrt{c_3}\right) g_h^3\right)}\),

\noindent\({\xi _6=216\ 2^{1/3} \pi ^{9/2} c_1^{1/3} \left(15 j^2 \sqrt{\pi }+28 g_h^3\right){}^2 \left(960 j^2 \sqrt{\pi }+\left(2048+15
	\sqrt{c_3}\right) g_h^3\right)}\),

\noindent\({\xi _7=43200 j^4 \pi ^4+5 \pi ^{3/2} \left(38400 j^2 \pi ^2+180 j^2 \pi ^2 \sqrt{c_3}-c_3^{3/2}\right) g_h^3}\),

\noindent\({\xi _8=3 \pi ^3 \left(70656+375 j^2 \pi ^2+560 \sqrt{c_3}\right) g_h^6+2400 \pi ^{9/2} g_h^9}\),

\noindent\({\xi _9=5\ 2^{2/3} \pi ^{3/2} c_1^{2/3} g_h^3+360\ 2^{1/3} \pi ^{9/2} g_h^3 \left(15 j^2 \sqrt{\pi }+28 g_h^3\right){}^2}\),

\noindent\({\xi _{10}=c_1^{1/3} \left(-32 c_3^{3/2}-480 \pi ^3 g_h^6+\sqrt{c_3} \left(5760 j^2 \pi ^2+12288 \pi ^{3/2} g_h^3-25 \pi ^3 g_h^6\right)\right)}\),

\noindent\({\xi _{11}=2160 \pi ^{9/2} c_1^{2/3} c_2 g_h^9 \left(256+25 \pi ^{3/2} g_h^3\right)}\),

\noindent\({c_1=-108 \left(c_{11}+\sqrt{c_{11}^2-16 \pi ^9 \left(15 j^2 \sqrt{\pi }+28 g_h^3\right){}^6}\right)}\),

\noindent\({c_{11}=\left(13500 j^6 \pi ^6+108000 j^4 \pi ^{11/2} g_h^3+273600 j^2 \pi ^5 g_h^6+223232 \pi ^{9/2} g_h^9-25 \pi ^6 g_h^{12}\right)}\),

\noindent\({c_2=\dfrac{2^{2/3} c_1^{2/3}+72\ 2^{1/3} \pi ^3 \left(15 j^2 \sqrt{\pi }+28 g_h^3\right){}^2+24 c_1^{1/3} \left(15 j^2 \pi ^2+32 \pi
		^{3/2} g_h^3\right)}{6 c_1^{1/3}}}\),

\noindent\({c_3=\dfrac{-120 \pi ^3 c_1^{1/3} g_h^6-\sqrt{c_2} \left(2^{2/3} c_1^{2/3}+72\ 2^{1/3} \pi ^3 \left(15 j^2 \sqrt{\pi }+28 g_h^3\right){}^2-48
		c_1^{1/3} \left(15 j^2 \pi ^2+32 \pi ^{3/2} g_h^3\right)\right)}{6 c_1^{1/3} \sqrt{c_2}}}\).

 Appendix C: Rotating Bardeen-AdS black holes with dark energy

\noindent\({P_{1S}=3180 b^8 \pi ^4-6360 b^4 j^2 \pi ^4-3180 b^6 \pi ^{7/2} \alpha +6360 b^6 \pi ^4 \alpha -9540 b^6 \pi ^{7/2} \alpha  \omega
	+9540 b^6 \pi ^4 \alpha  \omega  }\),\

\noindent\({{P_{2S}=-2610 b^2 j^2 \pi ^3 S_c-1305 b^4 \pi ^{5/2} \alpha  S_c+3480 b^4 \pi ^3 \alpha  S_c-3915 b^4 \pi ^{5/2} \alpha  \omega  S_c+5220 b^4 \pi ^3 \alpha  \omega
		S_c-156 b^4 \pi ^2 S_c^2}}\),\

\noindent\({P_{3S}= -540 j^2 \pi ^2 S_c^2-270 b^2 \pi ^{3/2} \alpha  S_c^2+876 b^2 \pi ^2 \alpha  S_c^2-810 b^2 \pi ^{3/2} \alpha  \omega  S_c^2+1314 b^2 \pi ^2
	\alpha  \omega  S_c^2-83 b^2 \pi  S_c^3}\),

\noindent\({P_{4S}= \left.+ 72 \pi \alpha  S_c^3+108 \pi  \alpha  \omega  S_c^3+6 S_c^4\right)}\),\

\noindent\({\lambda _5=\left(12 \lambda _3^{1/3} \left(358 b^6 \pi ^3+b^2 \pi ^{3/2} \left(870 j^2 \pi ^{3/2}+\alpha  \left(-15 (1+3 \omega )+43
	\sqrt{\pi } (2+3 \omega )\right) (9 \pi  \alpha  (2+3 \omega )-2 \right.\right.\right.}\\
{\left.\left.\sqrt{\lambda _4}\right)\right)+b^4 \pi ^2 \left(3 \sqrt{\pi } \alpha  \left(145 (1+3 \omega )-358 \sqrt{\pi } (2+3 \omega )\right)+55
	\sqrt{\lambda _4}\right)-3 \pi ^2 \left(9 \pi  \alpha  (2+3 \omega ) \left(10 j^2+\right.\right.}\\
{\left.\left.\left.\alpha ^2 (2+3 \omega )^2\right)-\left(20 j^2+3 \alpha ^2 (2+3 \omega )^2\right) \sqrt{\lambda _4}\right)\right)-18\ 2^{1/3}
	\pi ^3 \left(17 b^4 \sqrt{\pi }-30 j^2 \sqrt{\pi }+b^2 \alpha  (-15 (1+3 \omega )\right.}\\
{\left.\left.\left.+14 \sqrt{\pi } (2+3 \omega )\right)\right)^2 \sqrt{\lambda _4}-2^{2/3} \lambda _3^{2/3} \sqrt{\lambda _4}\right)/\left(6
	\lambda _3^{1/3} \sqrt{\lambda _4}\right)}\),\

\noindent\({\lambda _4=\left(18\ 2^{1/3} \pi ^3 \left(17 b^4 \sqrt{\pi }-30 j^2 \sqrt{\pi }+b^2 \alpha  \left(-15 (1+3 \omega )+14 \sqrt{\pi
	} (2+3 \omega )\right)\right)^2+6 \left(55 b^4 \pi ^2+2 b^2\right.\right.}\\
{\left.\left. \pi ^{3/2} \alpha  \left(15+45 \omega -43 \sqrt{\pi } (2+3 \omega )\right)+3 \pi ^2 \left(20 j^2+3 \alpha ^2 (2+3 \omega )^2\right)\right)
	\lambda _3^{1/3}+2^{2/3} \lambda _3^{2/3}\right)/\left(6 \lambda _3^{1/3}\right)}\),\

\noindent\({\lambda _3=54 \pi ^{9/2} \lambda _1+\sqrt{\lambda _2}}\),\

\noindent\({\lambda _2=2916 \pi ^9 \left(-\left(17 b^4 \sqrt{\pi }-30 j^2 \sqrt{\pi }+b^2 \alpha  \left(-15 (1+3 \omega )+14 \sqrt{\pi } (2+3
	\omega )\right)\right)^6+\lambda _1^2\right)}\),\

\noindent\({\lambda _1=4737 b^{12} \pi ^{3/2}-27000 j^6 \pi ^{3/2}+2700 b^2 j^4 \pi  \alpha  \left(-15 (1+3 \omega )+14 \sqrt{\pi } (2+3 \omega
	)\right)+3 b^{10} \pi  \alpha  (-3155}\\
{\left. (1+3 \omega )+2574 \sqrt{\pi } (2+3 \omega )\right)+b^6 \alpha  \left(-180 j^2 \pi  \left(-355 (1+3 \omega )+414 \sqrt{\pi } (2+3 \omega
	)\right)+\alpha ^2 (-3375 \right.}\\
{\left.\left.(1+3 \omega )^3+9450 \sqrt{\pi } (1+3 \omega )^2 (2+3 \omega )-7920 \pi  (1+3 \omega ) (2+3 \omega )^2+1576 \pi ^{3/2} (2+3 \omega
	)^3\right)\right)-3 b^8 }\\
{\sqrt{\pi } \left(6310 j^2 \pi -\alpha ^2 \left(5325 (1+3 \omega )^2+7648 \pi  (2+3 \omega )^2-12420 \sqrt{\pi } \left(2+9 \omega +9
	\omega ^2\right)\right)\right)+90 b^4 j^2 \sqrt{\pi }}\\
{ \left(710 j^2 \pi -\alpha ^2 \left(225 (1+3 \omega )^2+176 \pi  (2+3 \omega )^2-420 \sqrt{\pi } \left(2+9 \omega +9 \omega
	^2\right)\right)\right)}\),\

\noindent\(P_{1L}= {3180 b^8 n^{1+3 \omega } \pi ^4-6360 b^4 j^2 n^{1+3 \omega } \pi ^4+6360 b^6 \pi ^4 \alpha +9540 b^6 \pi
	^4 \alpha  \omega -2610 b^2 j^2 n^{1+3 \omega } \pi ^3 S_c}\),

\noindent\({P_{2L}=	{3480 b^4 \pi ^3 \alpha  S_c+5220 b^4 \pi ^3 \alpha  \omega  S_c-156 b^4 n^{1+3 \omega } \pi ^2 S_c^2-540 j^2 n^{1+3 \omega } \pi ^2 S_c^2+876
		b^2 \pi ^2 \alpha  S_c^2 }}\),

\noindent\({P_{3L}={1314 b^2 \pi ^2\alpha  \omega  S_c^2-83 b^2 n^{1+3 \omega } \pi  S_c^3+72 \pi  \alpha  S_c^3+108 \pi  \alpha  \omega  S_c^3+6 n^{1+3 \omega } S_c^4}}\),

\noindent\({}\)

\noindent\({\beta _1=4737 b^{12} n^{3+9 \omega }-27000 j^6 n^{3+9 \omega }+7722 b^{10} n^{2+6 \omega } \alpha  (2+3 \omega )+37800 b^2 j^4 n^{2+6
		\omega } \alpha  (2+3 \omega )- n^{1+3 \omega } }\\
{6 b^8\left(3155 j^2 n^{2+6 \omega }-3824 \alpha ^2 (2+3 \omega )^2\right)+180 b^4 j^2 n^{1+3 \omega } \left(355 j^2 n^{2+6 \omega }-88 \alpha ^2
	(2+3 \omega )^2\right)+8 b^6 \alpha  (2+3 \omega ) }\\
{\left(-9315 j^2 n^{2+6 \omega }+197 \alpha ^2 (2+3 \omega )^2\right)}\),\

\noindent\({\beta _2=54 \pi ^6 \beta _1+54 \pi ^6 \sqrt{-\left(17 b^4 n^{1+3 \omega }-30 j^2 n^{1+3 \omega }+14 b^2 \alpha  (2+3 \omega )\right)^6+\beta
		_1^2}}\),\

\noindent\({\beta _3=n^{-2-6 \omega } \left(18\ 2^{1/3} n^{1+3 \omega } \pi ^4 \left(17 b^4 n^{1+3 \omega }-30 j^2 n^{1+3 \omega }+14 b^2 \alpha
	(2+3 \omega )\right)^2+6 \pi ^2 \left(55 b^4 n^{2+6 \omega }+ n^{2+6 \omega }\right.\right.}\\
{\left.\left.60 j^2 -86 b^2 n^{1+3 \omega } \alpha  (2+3 \omega )+9 \alpha ^2 (2+3 \omega )^2\right) \beta _2^{1/3}+2^{2/3} n^{1+3 \omega } \beta _2^{2/3}\right)/\left(6
	\beta _2^{1/3}\right)}\),\

\noindent\({\beta _4= \left(12 \pi ^2 \beta _2^{1/3} \left(358 b^6 n^{3+9 \omega } \pi +b^2 n^{1+3 \omega } \left(870 j^2 n^{2+6
		\omega } \pi +43 \alpha  (2+3 \omega ) \left(9 \pi  \alpha  (2+3 \omega )-2 n^{1+3 \omega } \sqrt{\beta _3}\right)\right)\right.\right.}\\
{+b^4 n^{2+6 \omega } \left(-1074 \pi  \alpha  (2+3 \omega )+55 n^{1+3 \omega } \sqrt{\beta _3}\right)+3 \left(-9 \pi  \alpha  (2+3 \omega )
	\left(10 j^2 n^{2+6 \omega }+\alpha ^2 (2+3 \omega )^2\right)+\right.}\\
{\left.\left.n^{1+3 \omega } \left(20 j^2 n^{2+6 \omega }+3 \alpha ^2 (2+3 \omega )^2\right) \sqrt{\beta _3}\right)\right)-18\ 2^{1/3} n^{2+6
		\omega } \pi ^4 \left(17 b^4 n^{1+3 \omega }-30 j^2 n^{1+3 \omega }+14 b^2 \alpha\right.}\\
{\left.\left.  (2+3 \omega )\right)^2 \sqrt{\beta _3}-2^{2/3} n^{2+6 \omega } \beta _2^{2/3} \sqrt{\beta _3}\right)n^{-3-9 \omega }/\left(6 \beta _2^{1/3}
	\sqrt{\beta _3}\right)}\),\

\noindent\({T_1=\left(S^{\frac{5}{2}+\frac{3 \omega }{2}}+8 P S^{\frac{7}{2}+\frac{3 \omega }{2}}+\pi ^{\frac{3 (1+\omega )}{2}} S \alpha  (2+3
	\omega )+b^2 \left(-2 \pi  S^{\frac{3 (1+\omega )}{2}}+\pi ^{\frac{5}{2}+\frac{3 \omega }{2}} \alpha  (5+3 \omega )\right)\right)}\),\

\noindent\({T_2=\left(9 S^{\frac{3 (1+\omega )}{2}}+48 P S^{\frac{5}{2}+\frac{3 \omega }{2}}+64 P^2 S^{\frac{7}{2}+\frac{3 \omega }{2}}-3 \pi
	^{\frac{3 (1+\omega )}{2}} \alpha  (2+3 \omega )^2-8 P \pi ^{\frac{3 (1+\omega )}{2}} S \alpha  \right.}
{\left.\left(8+18 \omega +9 \omega ^2\right)\right)}\),\

\noindent\({T_3=\left(-45 S^{\frac{3 (1+\omega )}{2}}-192 P S^{\frac{5}{2}+\frac{3 \omega }{2}}-192 P^2 S^{\frac{7}{2}+\frac{3 \omega }{2}}+3
	\pi ^{\frac{3 (1+\omega )}{2}} \alpha  \left(20+33 \omega +18 \omega ^2\right)+8 P \pi ^{\frac{3 (1+\omega )}{2}}\right.}\\
{\left. S \alpha  \left(34+45 \omega +18 \omega ^2\right)\right)}\),\

\noindent\({T_4=\left(-6 \pi ^2 S^{\frac{3 (1+\omega )}{2}} (3+8 P S)+\pi ^{\frac{7}{2}+\frac{3 \omega }{2}} \alpha  (5+3 \omega ) (6+9 \omega
	+8 P S (4+3 \omega ))\right)}\),\

\noindent\({T_5=\left(2 \left(b^2 \pi +S\right)^5 \left(3 S^{\frac{3 (1+\omega )}{2}}+8 P S^{\frac{5}{2}+\frac{3 \omega }{2}}-3 \pi ^{\frac{3
			(1+\omega )}{2}} \alpha \right)^2\right)}\),\

\noindent\({M_1=6 S \left(b^2 \pi +S\right) \left(3 S^{\frac{3 (1+\omega )}{2}}+8 P S^{\frac{5}{2}+\frac{3 \omega }{2}}-3 \pi ^{\frac{3 (1+\omega
			)}{2}} \alpha \right)}\),\

\noindent\({M_2=\left(9 S^{\frac{3 (1+\omega )}{2}}+48 P S^{\frac{5}{2}+\frac{3 \omega }{2}}+64 P^2 S^{\frac{7}{2}+\frac{3 \omega }{2}}-72 P
	\pi ^{\frac{3 (1+\omega )}{2}} S \alpha  (2+\omega )-9 \pi ^{\frac{3 (1+\omega )}{2}} \alpha  (2+3 \omega )\right)}\),\

\noindent\({M_3=b^2 \pi  \left(36 S^{\frac{3 (1+\omega )}{2}}+192 P S^{\frac{5}{2}+\frac{3 \omega }{2}}+256 P^2 S^{\frac{7}{2}+\frac{3 \omega
		}{2}}-72 P \pi ^{\frac{3 (1+\omega )}{2}} S \alpha  (3+\omega )-9 \pi ^{\frac{3 (1+\omega )}{2}} \alpha  (5+3 \omega )\right)}\),\

\noindent\({M_4=\left(3 S^{\frac{3 (1+\omega )}{2}}+8 P S^{\frac{5}{2}+\frac{3 \omega }{2}}-3 \pi ^{\frac{3 (1+\omega )}{2}} \alpha \right)^2}\).

\end{document}